\documentclass[final,4p]{elsarticle}
\usepackage{amssymb}
\usepackage{graphicx}
\usepackage{multirow}
\usepackage{txfonts}
\begin{document}

\begin{frontmatter}

\title{Monte Carlo simulation to investigate the formation of molecular
hydrogen and its deuterated forms}

\author[ICSP]{Dipen Sahu}
\ead{dipen@csp.res.in}
\author[ICSP]{Ankan Das}
\ead{ankan.das@gmail.com}
\author[ICSP]{Liton Majumdar}
\ead{liton@csp.res.in}
\author[SNBNCBS,ICSP]{Sandip K. Chakrabarti}
\ead{chakraba@bose.res.in}
\address[ICSP]{Indian Centre For Space Physics, 43 Chalantika, Garia Station Road, Kolkata 700084, India}
\address[SNBNCBS]{S.N. Bose National Center for Basic Sciences, JD-Block, Salt Lake, Kolkata,700098, India}

\begin{abstract}
$H_2$ is the most abundant interstellar species. Its deuterated forms ($HD$ and $D_2$) 
are also significantly abundant. Huge abundances of these molecules could be
explained by considering the chemistry occurring on the interstellar dust.
Because of its simplicity, Rate equation method is widely used to study the formation of 
grain-surface species. However, since recombination efficiency of formation of
any surface species are heavily dependent on various physical and 
chemical parameters, Monte Carlo method  would be best method suited to 
take care of randomness of the processes. We perform Monte Carlo simulation to 
study the formation of $H_2$, $HD$ and $D_2$ on interstellar ices. 
Adsorption energies of surface species are the key inputs for the formation of 
any species on interstellar dusts but binding energies of
deuterated species are yet to known with certainty. A zero point energy 
correction exists between hydrogenated and deuterated species which should be considered
while modeling the chemistry on the interstellar dusts. Following some earlier studies,
we consider various sets of adsorption energies to study the formation of
these species in diverse physical circumstances. As expected, noticeable difference
in these two approaches (Rate equation method and Monte Carlo method)
is observed for production of these simple molecules 
on interstellar ices. We introduce two factors, namely, $S_f$ and $\beta$ 
to explain these discrepancies: $S_f$ is a scaling factor, which could be used to 
correlate discrepancies between Rate equation and Monte Carlo methods. $\beta$ 
factor indicates the formation efficiency under various circumstances. 
Higher values of $\beta$ indicates a lower production efficiency. 
We found that $\beta$ increases with a decrease in rate of accretion from gas phase to grain phase.
\end{abstract}


\end{frontmatter}

\section{Introduction}
Molecular hydrogen is the most abundant and simplest species in the 
Interstellar Medium (ISM). Indeed, $H_2$ is the most important molecule, as it is the 
first and foremost precursor to create other complex molecules. 
Simpler molecules are mainly formed on grain surfaces and then desorbed 
to gas phase (Gould \& Salpeter, 1963). Importance of grain chemistry 
were already described in the earlier literatures 
(Hasegawa, Herbst \& Leung, 1992; Chakrabarti et. al., 2006ab; Cuppen \& Herbst, 2007; 
Das et al., 2008ab; Das et al., 2010, 2013ab; Das \& Chakrabarti, 2011; Majumdar et al., 2012; 2013; 
Das et al. 2014). Sometimes even without explicit grain chemistry, effective  rates were used 
in studying  the formation of bio-molecules (Chakrabarti \& Chakrabarti, 2000ab) 
For large grains or high accretion rates, Rate equation method 
(Biham et al., 2001) can be used to explain abundances of surface species. But due to inherent 
randomness of the processes, especially when grain size is small and/or accretion rate is small, 
simple Rate equation method is not good enough to explain recombination efficiency of 
interstellar surface species. It is then essential to consider Monte Carlo method 
to account all the features.  

Despite low elemental abundances of atomic deuterium (having  D/H ratio of $\sim 10^{-5}$ 
according to Linsky et al., 1995), several complex molecules are found to be heavily 
fractionated (Majumdar, Das \& Chakrabarti, 2014ab) in ISM. 
As like the normal hydrogen molecules, its deuterated forms could also be synthesized 
on interstellar grains. Chakrabarti et al. (2006ab) carried 
out Monte Carlo simulation to find out exact recombination efficiency for the formation of $H_2$.
Following similar approach, here we  use both Rate equation method and Monte Carlo 
method to find out the various aspects during the formation of  $H_2$, $HD$ and $D_2$. 

The plan of this paper is the following. In Section 2, computational details are presented. 
Implications of results are discussed in Section 3. Finally, in Section 4, we draw our conclusions.

\section{Computational details}

\subsection {Rate Equation Method:} 
A small chemical network containing $H$ and $D$ is used here.
Let at any time $t$, $N_x$ be the number of species $x$ on a grain having $S$ number of adsorption
sites. Governing equations for all surface species in our network could then be written as follows:

\begin{equation}
\frac {d N_H}{dt}= F_H- W_H N_H -2 a_H N_H^2/S - a_{HD}N_H N_D/S, 
\end{equation}

\begin{equation}
\frac {dN_{H_2}}{dt}=F_{H_2}+ \mu_{H_2} a_H N_H^2/S- W_{H_2} N_{H_2},    
\end{equation}

\begin{equation}
\frac{dN_{D_2}}{dt}=F_{D_2}+\mu_{D_2} a_D N_D^2/S - W_{D_2} N_{D_2},  
\end{equation}

\begin{equation}
\frac{dN_D}{dt}=F_D-W_D N_D-2 a_D N_D^2/S- a_{HD} N_H N_D/S,
\end{equation}

\begin{equation}
\frac{dN_{HD}}{dt}=F_{HD} + \mu_{HD} a_{HD}N_H N_D/S - W_{HD}N_{HD},    
\end{equation}
where, $F_x$, $a_x$ and $W_x$ represents accretion, hopping and desorption
rates respectively for species $x$ in the units of $sec^{-1}$.
$\mu_x$ represents spontaneous desorption factor. It is assumed that $(1-\mu_x)$ factor of 
$x^{th}$ species could go to gas phase just after its formation on grain. 
From experimental findings of Katz et al. (1999), we use $\mu_{H_2}=0.33$ for olivine 
grains and $\mu_{H_2}=0.413$ for amorphous carbon grains. Here, we use similar spontaneous
desorption factor for $HD$ and $D_2$ as well. Accretion rate (F$_x$) of $x^{th}$ species 
is calculated from,
\begin{equation}
F_x= t_x A\left <V \right> N_x,
\end{equation}
where, $t_x$ is the sticking coefficient, $A$ is the surface area of the grain in the units of
$cm^2$, $<V>$ is average thermal velocity ($<V>= \sqrt{\frac{8kT}{m\pi}} \ cm \ sec^{-1}$) and
$n_x$ is the gas phase concentration of $x^{th}$ species (in $cm^{-3}$).

\subsubsection{Sticking coefficient}

Since the sticking coefficient of $H_2$ is $\sim 0$ at $T=10 \ K$ (Leith \& Williams, 1985),
here, We consider, sticking coefficients ($t_x$) of $H_2$, $HD$ and $D_2$ to be $\sim 0$. 
Thus, from Eqn. 6, we have $F_{H_2}, F_{HD}$ \& $F_{D_2}=0$. 
In reality, sticking probability of a species is mainly governed by kinetic energy of 
the incoming species and adsorption energy of that species with  grain surface.
Since, hydrogen is the lighter species, hydrogenation reaction is the
fastest reaction on the grain surface. In order to determine the sticking coefficient,  
various attempts were made over the last few decades. Buch \& Zhang (1991) 
performed molecular dynamical simulation to numerically evaluate sticking 
of hydrogen atoms with cluster of water molecules. According to their study, 
sticking coefficient depends on the following:
\begin{equation}
S=(K_BT/E_0+1)^{-2},
\end{equation}
where, $E_0/K_B=102 \ K$ for $H$ atom and $200 \ K$ for $D$ atom. As per Chaabouni et al. (2012)
sticking parameter of $\sqrt{10/T}$ could be adopted. Recently, Matar et al. (2010) performed an 
experiment to model the sticking parameters of $H_2$ and $D_2$  and its dependence on 
impinging molecular beam temperature. They found out an analytic formula (Eqn. 8) to describe sticking 
coefficient at any temperature $T$. From the outcome of their experiments and the
interpretation of Chaabouni et al. (2012), sticking coefficient could be parameterized 
for hydrogen and deuterium on silicate surfaces under interstellar conditions. 
They pointed out that sticking coefficients of these species with silicate grains behave
in the same as they would on icy dust grains. Their prescribed variation of sticking parameter 
is as follows:
\begin{equation}
S(T)=S_0 \frac{(1+\beta T/T_0)}{{(1+T/T_0)}^{\sigma}},
\end{equation}
where, $S_0$ is sticking coefficient of particles at zero temperature and $T_0$
is a critical temperature obtainable using experimental data. 
$\sigma$ defines the geometry of incident beam. Here, in our simulation, unless otherwise stated
we always consider the sticking coefficient of all the species $1$. For a special case, we 
consider the assumption of Chaabouni et al. (2012).

\subsubsection{Accretion, Diffusion and Desorption}

An effective accretion rate on a typical grain is defined as, 
$\phi_x=F_x(1-\Sigma_i {f_{gr}}_{xi})$, where, $\Sigma_i {f_{gr}}_{xi}$ is 
fraction of grain that could be occupied by $x_i$ number of species. Hopping rate of a 
species $x$ is calculated by,
\begin{equation}
a_x=\nu_x \exp^{-{E_b}_x/K_bT},
\end {equation} 
where, $\nu_x$ is vibrational frequency and ${E_b}_x$ is energy barrier for diffusion. 
Vibrational frequency of any species $x$ is calculated by,
\begin{equation}
\nu_x=\sqrt{\frac{2s{E_d}_x}{\pi ^2 m_x}},
\end{equation}
where, $m_x$ is mass of species $x$, $E_d$ is adsorption energy of species $x$,  
$s$ is surface density of sites in the unit of $cm^{-2}$ ($s=10^{14}$ cm$^{-2}$ is used).
Finally, desorption rate of a species $x$ is calculated by,
\begin{equation}
W_x=\nu_x \exp(-{E_b}_x/K_b T).
\end{equation}

\subsubsection{Binding energies}

Binding energies are important to decide the mobility of surface species.
From experimental findings of Pirronello et al. (1997, 1999), Katz et al. (1999)  
concluded that atomic hydrogen moves much more slowly than what is
normally used in various simulations. In our simulation, we use these experimental 
findings. In Katz et al. (1999), $E_b$ and $E_d$ were obtained for $H$ atom. 
However, for $H_2$ molecules, only $E_d$ was defined.  
Since interaction between adsorbed $H$ and $D$ atoms and grains are different,
a zero point energy correction due to isotopic effects should to be considered
because of which binding energies of deuterated species must differ. 
$D$ atom being heavier than $H$ atom, it would sit at a level lower than 
$H$ in any potential surface if zero-point energy is included (Lipschitz, Biham \& Herbst, 2004).
Caselli et al. (2002) proposed a difference of $2 \ meV$ between desorption energies and 
barriers against diffusion. Following this assumption, Lipshtat, Biham \& Herbst (2004) (hereafter LBH) 
considered a $2 \ meV$ energy difference between $H$ and $D$ atoms for barriers against diffusion. 
They considered energy barriers against diffusion for $H$ atoms to be $35 \ meV$ whereas 
for $D$ atoms they considered it is to be $37 \ meV$. In case of desorption energies, 
they considered for $H$ atoms it is $50 \ meV$ and for $D$ atoms it is $60 \ meV$. 
So, in case of barrier against desorption,  LBH considered an energy difference of $10 \ meV$.  
As per LBH, this larger difference in case of desorption was considered because barrier against 
diffusion balances zero-point energy of a potential well with that of a saddle point 
while desorption does not possess a saddle point.

In absence of a consensus on binding energies,
we use three sets of energy barriers in our simulations. These sets are 
shown in Table 1. First set corresponds to experimental binding energies which 
were taken from Katz et al. (1999) for Olivine grains. Second set corresponds to 
experimental values obtained for Amorphous carbon grains (Katz et al., 1999) and third 
set is same as LBH. Binding energies for deuterated species are calculated by using 
following scaling relations,
$$
{\scriptsize
E_b(D)=E_b(H) \times \frac{37 \ (E_b(D) \ from \ LBH)}{35 \ (E_b(H) \ from \ LBH)},}
$$
$$
{\scriptsize
E_d(D)=E_d(H) \times \frac{60 \ (E_d(D) \ from \ LBH)}{50 \ (E_d(H) \ from \ LBH)}},
$$
$$
{\scriptsize
E_d(HD,D_2)=E_d(D) \times \frac{46.7 \ (E_d(H_2) \ for \ amorphous \ carbon \ from \ Katz \ et \ al. \ (1999))}
{56.7 \ (E_d(H) \ for \ amorphous \ carbon \ from \ Katz \ et \ al. \ (1999))}}
$$
Following Hasegawa, Herbst \& Leung (1992); Das et al. (2008) and 
references therein, we consider $E_d (x)=0.3 \ E_b (x)$ for $H_2$, $HD$ and $D_2$ molecules.
In case of set 3, desorption energy for $H_2$ was not available. 
We use following relation to compute desorption energy of $H_2$ for set 3:
$$
{\scriptsize
E_d(H_2)=E_d(H) \times \frac{46.7 \ (E_d(H_2) \ for \ amorphous \ carbon \ from \ Katz \ et \ al. \ (1999))}
{56.7 \ (E_d(H) \ for \ amorphous \ carbon \ from \ Katz \ et \ al. \ (1999))}}
$$

\begin{table}
{\centering
\scriptsize
\caption{Binding energies used in our simulations}
\begin{tabular}{|c|c|c|c|c|c|c|}
\hline
Species&\multicolumn{2}{|c|}{Olivine grain}&\multicolumn{2}{|c|}{Amorphous carbon grain}
&\multicolumn{2}{|c|}{Lipshtat, Biham \& Herbst (2004)}\\
&\multicolumn{2}{|c|}{(set 1)}&\multicolumn{2}{|c|}{(set 2)}
&\multicolumn{2}{|c|}{(set 3)}\\
\hline
&E$_b$ (in meV)&$E_d$ (in meV)&E$_b$ (in meV)&E$_d$ (in meV)&E$_b$ (in meV)&E$_d$ (in meV)\\
\hline\hline
H&24.7&32.1&44.0&56.7&35.0&50.0\\
D&26.1&38.5&46.5&68.0&37.0&60.0\\
H$_2$&8.1&27.1&14.0&46.7&12.4&41.2\\
HD&9.8&32.5&16.8&56.0&14.8&49.4\\
D$_2$&9.8&32.5&16.8&56.0&14.8&49.4\\
\hline
\end{tabular}}
\end{table}

In reality, nature of grain surface need not be simple. There might be various types of lattice defects
and various phases. Defects with enhanced binding energies (Hollenbach \& Salpeter, 1971)
could allow formation of $H_2$ up to $50\ K$. In our simulation, we do not consider such types of grains
and use a very low temperature window ($8\ K-25\ K$).

\subsection {Procedures adopted in a Monte Carlo Simulation}

In order to preserve randomness in molecular formation process, it is essential 
to develop Monte Carlo algorithm to mimic exact scenario. 
Formation of molecules on grains are mainly due to two types of forces, 
namely, (i) the weak Van der Walls interaction and (ii) strong covalent 
bond (chemisorption). For low temperature region and/or for 
low surface temperature, atoms are weakly bound to surfaces via first 
type of force called physisorption. Here energy is of the order 
of a few $meV$. As species are weakly bound to surfaces, they can 
move around the surfaces via quantum mechanical tunneling or by thermal 
hopping. These processes are completely random. A species can 
move in any directions on the grain surface and could combine with another species
to form a third species. We assume the surface to be a square lattice having $n \times n $ 
number of sites in each direction. To mimic spherical nature of 
the grain, we consider periodic boundary condition. Simulation on a complex shaped grain is 
outside the scope of the present work, and will be taken up elsewhere.

Governing equations (Eqn. 1-5) are implemented in Monte Carlo method by assuming a four step process. 
First step is accretion of gas phase species onto a grain surface. Second step is diffusion 
of surface species. Third step is reaction among hopping species (i.e., at chance-meeting site).
Finally, fourth step is evaporation of the product to gas phase. Reaction could 
occur via thermal diffusion or quantum mechanical tunneling. Since, we are 
considering experimental values of binding energies, only thermal hopping is 
considered. Production of surface species via Eley-Rideal mechanism (Das et al., 2008 
and references therein) are also considered. All the possibilities in Monte Carlo 
method are handled by generating the random numbers as and when required.

In our simulation, smallest time scale is the hopping time scale of 
$H$ or $D$ atoms. Accretion time scale is much longer than hopping time scale.
Therefore, for simplicity, we are dropping a hydrogen atom 
after every $a_H/F_H$ steps and a deuterium atom in every $a_D/F_D$ steps. 
Dropping locations are automatically chosen by generating a pair of random numbers. 
Depending on the number ($n$) of sites in each direction of the grain concerned, 
random numbers are scaled. While dropping $H$ or $D$, if the designated location 
is already occupied by $H_2$, $HD$ or $D_2$ then we look for a different site for 
its landing. However, if the site is  occupied by $D$ or $H$ then $HD/D_2$ or $H_2/HD$ 
could be formed. Depending on binding energies, surface species are allowed 
to hopp towards any of the four directions. Directions are chosen 
by generating random numbers. When an atom meets another atom via 
thermal hopping, a new species could be formed with certainty 
(i.e., no activation barrier is considered). Evaporation of surface species are also handled 
by generating random numbers. Desorption rate for any surface species is calculated by
using Eqn. 9 which implies that species $x$ will leave after $1/W_x$ seconds. 
We are generating random numbers after every $1/a_H$ seconds to check this 
desorption probabilities. If this number is less than $W_x/a_x$ 
then selected $x$ species are released from grain phase. There are another type of desorption, namely 
spontaneous desorption, which are also handled by generating random numbers. This 
desorption term decides whether any newly formed species will remain on grain surface
or desorbed from the surface.

\section{Results and Discussion}

In Fig. 1, chemical evolution of the surface species ($\mathrm{H,\ H_2,\ D,\ HD \ \& \ D_2}$) 
are shown. Results from both Rate equation and Monte Carlo method are shown. Clear 
difference between these two methods are visible. Here, we consider amorphous carbon grains 
having hydrogen (atomic form) number density ($n_H) = 10^2 \ cm^{-3}$, $T=15 \ K$ and initial atomic
$D/H$ ratio (hereafter $r_D$) $= 0.1$. Surface abundance of $H$ atom has achieved a steady state after
$ \sim 10^5$ seconds, whereas, for $D$ atoms, steady state is achieved after $ \sim 10^6$ seconds.
For this set of physical parameters, $HD$ is the most abundant species on grain surface.
A steady state would be achieved beyond $2 \times 10^7$ seconds. It is to be noted 
for this choice of physical parameters $HD$ and $D_2$ are overproduced in 
the Rate equation method (Fig. 1).

\begin {figure}
\centering{
\includegraphics[height=10cm,width=10cm]{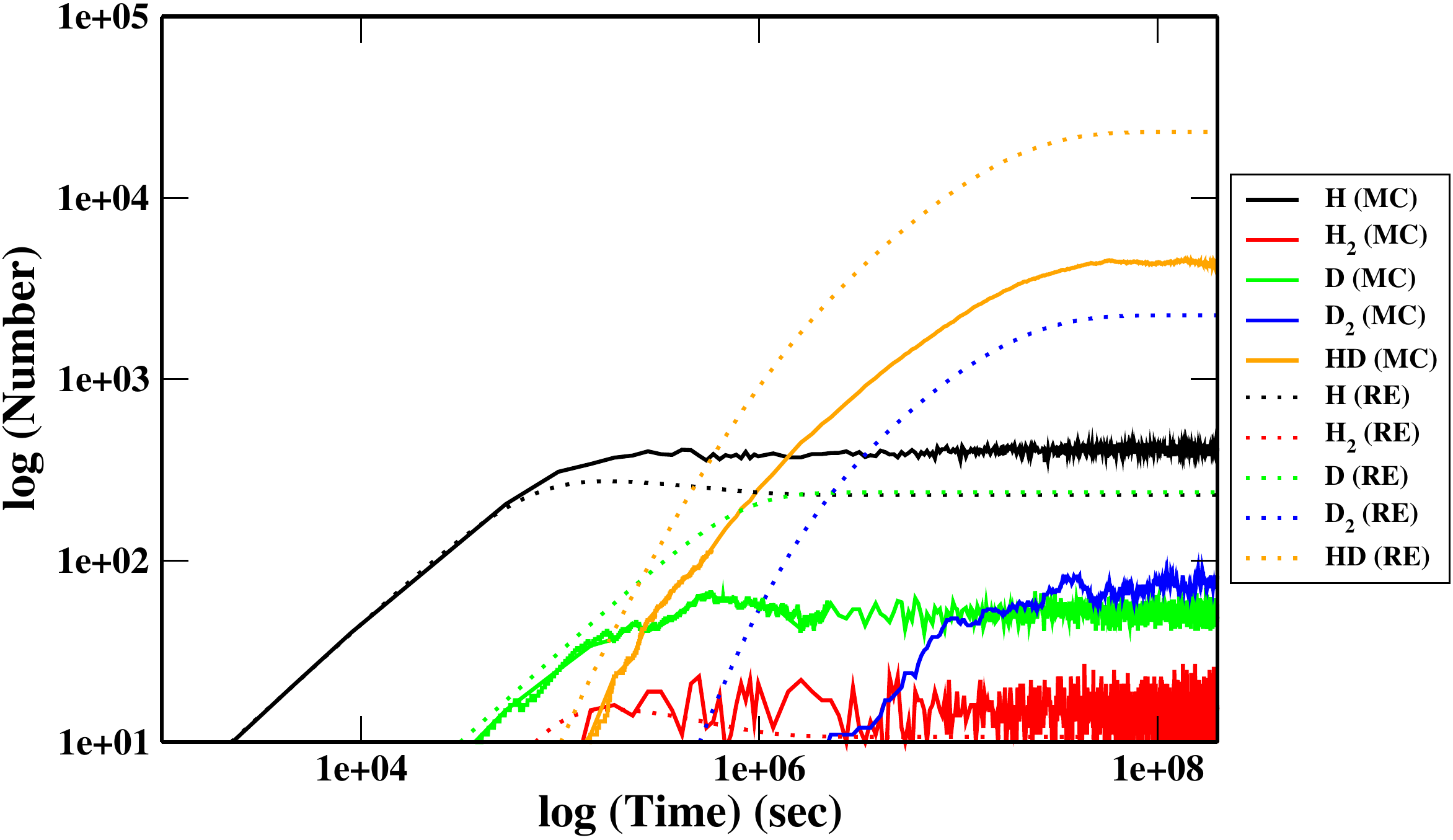}
\caption{Time evolution of abundances of $\mathrm{H,\ H_2,\ D,\ HD \ \& \ D_2}$ 
by Rate equation method (dotted lines) and Monte Carlo method (solid lines).
Simulation were carried out for an amorphous carbon grain kept at $15\ K$,
$n_H=10^2$ cm$^{-3}$ and $r_D=0.1$. }
\label{fig-1}}
\end {figure}

\begin {figure}
\centering{
\includegraphics[height=9cm,width=15cm]{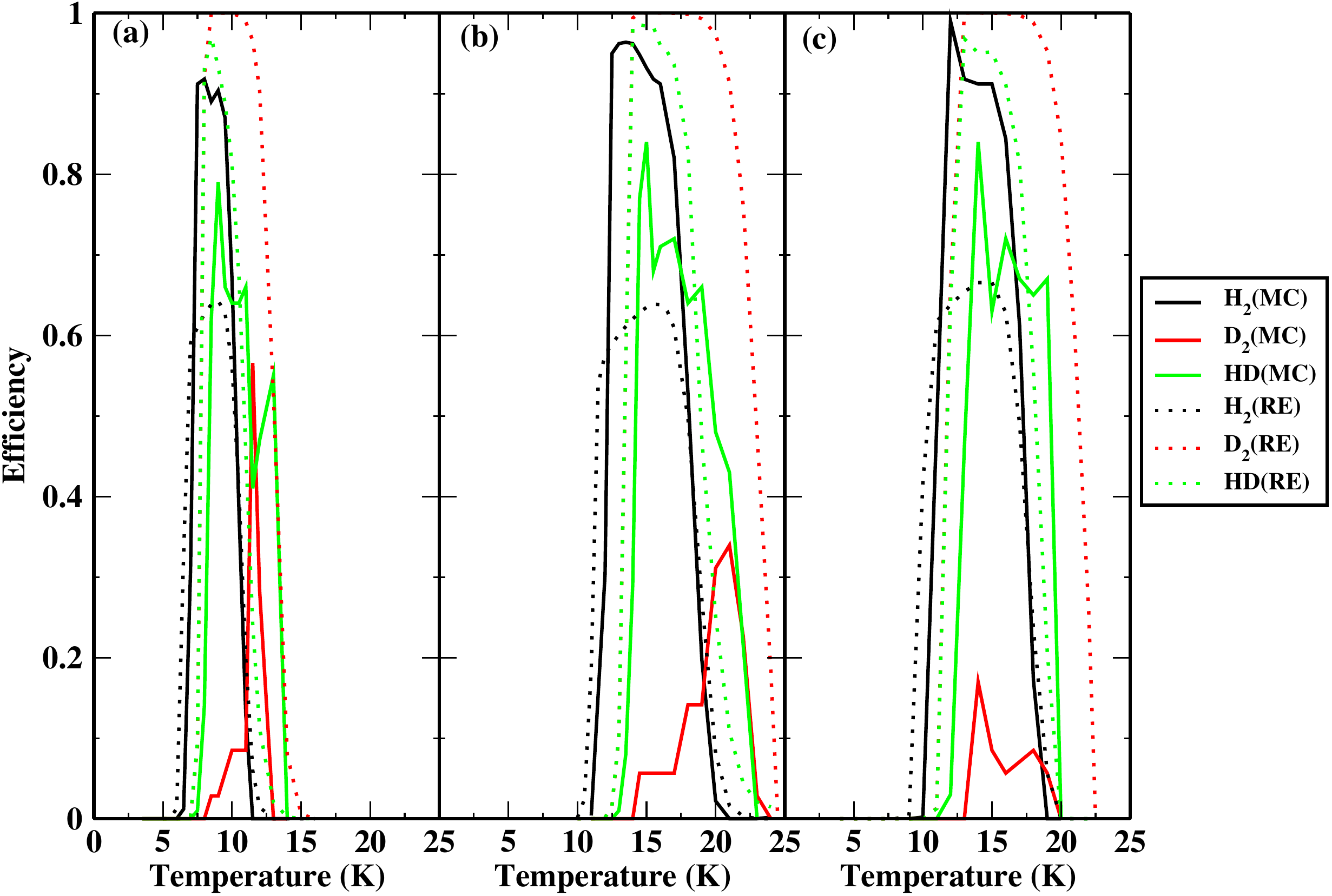}
\caption{Efficiency window for various sets of binding energies for (a) Olivine, (b)
Amorphous Carbon and (c)  for LBH case. Solid lines are results obtained from Monte
Carlo method whereas dotted lines are results from Rate equation method.}
\label{fig-2}}
\end {figure}

In Fig. 2, efficiency window for the formation of $H_2$, $D_2$ and $HD$ molecules are shown 
for Olivine grain (Fig. 2a), Amorphous carbon grain (Fig. 2b) and a type of grain having 
binding energies intermediate of Olivine and Amorphous Carbon type (Fig. 2c, with set 3 energy
values of Table 1) 
(Fig. 2c) are shown for $n_H=10^2 \ cm^{-3}$ and $r_D=0.1$. Production efficiency is defined by 
following relation:
\begin{equation}
\eta_{H_2}= \frac{2 R_{H_2}}{F_H},
\end{equation}
\begin{equation}
\eta_{D_2}= \frac{2 R_{D_2}}{F_D},
\end{equation}
\begin{equation}
\eta_{HD}= \frac{R_{HD}}{F_H+F_D}.
\end{equation}
where, $R_{H_2},\ R_{D_2}$ and $R_{HD}$ are gas phase productions of $H_2, \ D_2$ and $HD$ 
respectively. Gas phase production of these species solely depends on barrier against 
desorption (i.e., thermal desorption). Thus,
$$
R_{H_2}=W_{H_2} N_{H_2},
$$
$$
R_{D_2}=W_{D_2} N_{D_2},
$$
$$
R_{HD}=W_{HD} N_{HD}.
$$
From, Fig. 2, distinct features are observed for various isotopes of $H_2$. 
Solid lines represents the results obtained from Monte Carlo method whereas
dotted lines are for the results of Rate equation method. Difference between 
these two methods could be understood from Fig. 2. 
Monte Carlo method gives exact production rates due to the consideration of randomness. 
Here, we are computing the efficiency by considering the average production 
during the last few seconds. We find an interesting feature in the results of Monte Carlo method
(Fig. 2abc). We have seen a dip in the efficiency profile of $HD$ at the location where
efficiency profile of $D_2$ having a peak. This feature is prominent in case of the Olivine grain (Fig. 2a). 
A decreasing feature in the efficiency profile of $HD$ is visible at $T=11.5 \ K$ (Fig. 2a) 
whereas at the same location maximum efficiency is obtained for $D_2$.
This means that due to efficient production of $D_2$, some $D$ atoms are used up, 
which in turn decrease the production efficiency of $HD$. In case of Fig. 2b and 2c this feature 
is not well pronounced due to the lower production of $D_2$.

In all the cases, Rate equation method shows lower efficiency (under production) for the production of $H_2$ 
but higher efficiency (over production) in case of $HD$ and $D_2$. In this context, 
we can compare Fig. 1 and Fig. 2b since in both the cases, we use amorphous carbon grain. 
Fig. 2b, shows that the production efficiency of 
$HD$ and $D_2$ are higher in case of Rate equation method, which justifies the
over production of $HD$ and $D_2$ in Rate equation method (Fig. 1). 
Since Monte Carlo method is more accurate, we recommend that this method must 
be used for all surface reactions. In case of Olivine grains, efficiency window for 
$H_2$ is wider and extended from $8\ K$ to $10\ K$. 
For $HD$, this window falls in between $9-14\ K$. For $D_2$,
window is very narrow and peak is obtained at around $11.5\ K$. 
For amorphous Carbon grain (Fig. 2b), this window belongs to $11\ K-16\ K$, $15\ K-19\ K$, $21\ K-22\ K$ 
respectively for $H_2$, $D_2$ and $HD$.
From, Fig. 2c, efficiency window comes out to be at $11-16\ $K, $14-19\ $K and $14-17\ $K 
respectively for  $H_2$, $D_2$ and $HD$. 
\begin {figure}
\centering{
\includegraphics[height=10cm,width=10cm]{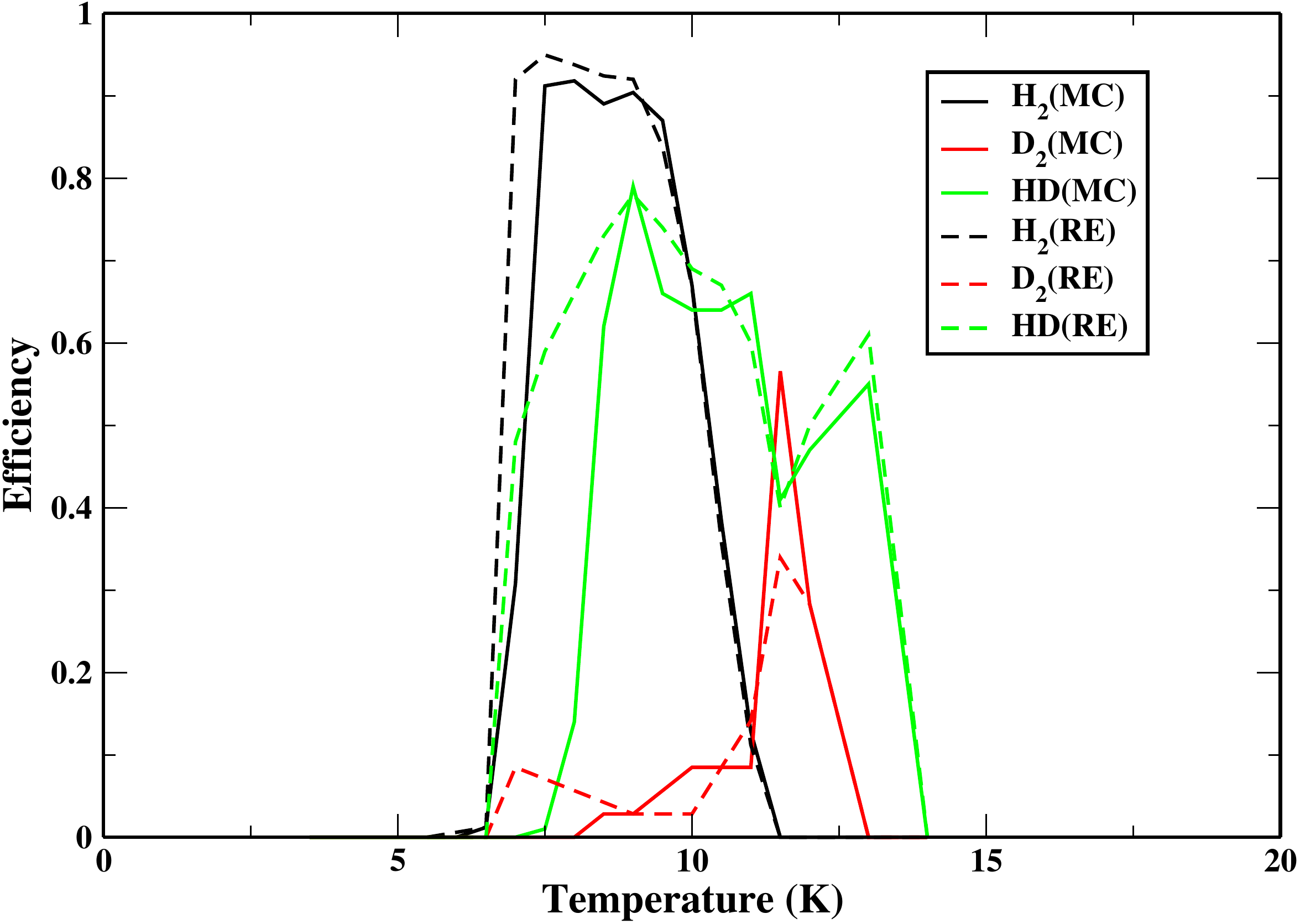}
\caption{Difference between cases when spontaneous desorption factor is included (dashed line) and not included (solid line). }
\label{fig-3}}
\end {figure}

So far, we discuss gas phase abundances of various isotopes of hydrogen molecules by
considering thermal desorption from interstellar grains only. However, there should be another 
probability of desorption by which gas phase would be populated. This is called spontaneous desorption 
process, where some fraction of species could be evaporated just after the formation. This happens
due to the energy liberated during some reactions. 
To account for this feature in our model, we consider that a factor 
$(1-\mu)$ of $N_{H_2}$, $N_{D_2}$ and $N_{HD}$ is lost to the gas phase. 
According to Katz et al. (1999), for an Olivine grain, $\mu_{H_2} = 0.33$ may 
be used and for amorphous carbon grain, $\mu_{H_2}= 0.413$ could be used. Due to 
unavailability of experimental data, here, we
use same $\mu$ values for $HD$ and $D_2$ ($\mu_{H_2}=\mu_{D_2}=\mu_{HD}$).
Here too, random numbers are generated for each newly formed $H_2, \ D_2$ and 
$HD$ and a fraction of species are allowed to populate the gas phase upon its production.
In Fig. 3, we compare the efficiency window of $H_2$, $D_2$ and $HD$
by considering the with (dashed line) and without spontaneous desorption term (solid line). 
Here we consider the Olivine grain for $n_H=10^2$ cm$^{-3}$ and $r_D=0.1$. 
While spontaneous desorptions are allowed, molecules are started to
produce efficiently in gas phase at somewhat lower temperatures. 
For example, production efficiency of $H_2$ attain a peak value
at $8\ K$ when spontaneous desorption term is not considered (solid line of Fig. 3). 
With spontaneous desorption factor (dashed line of Fig. 3)  this comes out to be at $7.5\ K$. 
For $HD$, solid line (no desorption) of Fig. 3 attain its maximum efficiency at 
$8.5\ K$ while dashed line (considering spontaneous desorption) also attain a peak
at $8.5 \ K$. But, while the spontaneous desorption term taken into account, $HD$ started to produce
at somewhat lower temperature. For example, a moderate efficiency at $7.5\ K$ is achieved
for the production of $HD$ molecule while spontaneous desorption term is considered but at $7.5\ K$, 
efficiency for the production of $HD$ is $\sim 0$ when spontaneous desorption term is not considered. 
The reason behind is that at low temperatures, thermal desorption time scales are much longer and 
spontaneous desorption factor allow some species to release from the grain surface and populate the gas 
phase. As a results, surface species starts to populate the gas phase little bit lower temperatures 
while considering the spontaneous desorption factor.

\begin {figure}
\centering{
\includegraphics[height=10cm,width=10cm]{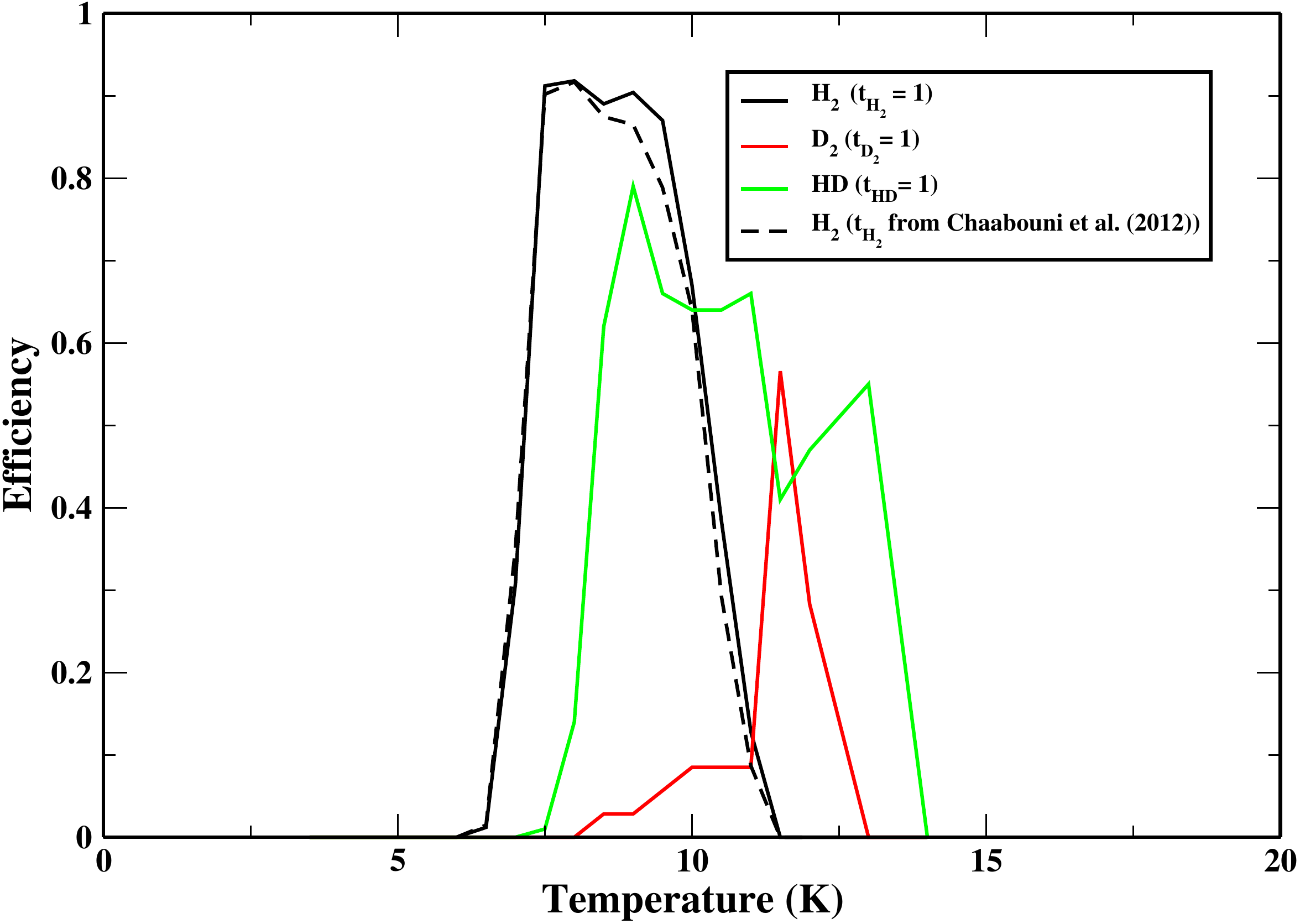}
\caption{Efficiency window of a Olivine grain by considering sticking coefficient 1 (solid lines) 
and sticking coefficient from Chaabouni et al. (2012) (dashed line).}
\label{fig-4}}
\end {figure}

Following Chaabouni et al. (2012), sticking coefficient of $H$ and $D$ are calculated by using Eqn. 8.
For a Silicate grain, Chaabouni et al. (2012) considered the sticking parameter ($S_0$) $=1$ for the both
$H$ and $D$ at $T_0 = 25\ K$ and $50\ K$.
From Eqn. 6, it is clear that the parameter is heavily dependent on temperature. 
In Fig. 4, a comparison between the consideration of unity sticking coefficient 
(normal case and used for all the cases) and the consideration by Chaabouni et al. (2012) 
is shown for an Olivine grain with $n_H=10^2$ cm$^{-3}$ and $r_D=0.1$. 
It is interesting to see that in the regime of 
our simulation, when we are considering sticking coefficients from Chaabouni et al. (2012), 
productions of $HD$ and $D_2$ are insignificant. Only significant amount of $H_2$ is produced in
this case. In between $5-13\ K$, sticking coefficient of $H$ as calculated from
Chaabouni et al., (2012) varies in the range of $0.95-0.80$. In case of $D$, it
varies in the range of $0.79-0.57$.

In Fig. 5(a-c), efficiency windows are shown when number of grain sites is varied.
Here we consider an Olivine grain kept inside $n_H=10^2$ cm$^{-3}$ and $r_D=0.1$. 
Fig. 5abc shows  efficiency window for Olivine grain having 
(a) $2.5 \times 10^2$ (Fig. 5a), (b) $4 \times 10^4$ (Fig. 5b) and (c) $1.6 \times 10^5$ (Fig. 5c)
number of sites respectively.
It is interesting to note that for larger grains, all molecules are producing efficiently. 
For smaller grains size (with accretion limit grain, having $\sim 2.5 \times 10^2$ sites), 
$D_2$ is not producing. However, for the larger grains ($1.6 \times 10^5$) it is producing.
From Fig. 5(a-c), it is clear that as we are increasing grain size, efficiency for the
formation of $D_2$ increases. 

It is discussed earlier that Monte Carlo method would provide a good estimation of 
recombination efficiency and Rate equation method often over or under estimate production
rate. To overcome these discrepancies, Chakrabarti et al. (2006ab) defined 
$A_x = a_x /S$, where $A_x$ is effective recombination rate of a surface species 
to recombine with another surface species and
$a_x$ is hopping rate as defined earlier. This is due to the fact that in two dimensions, 
a species (random walker) needs some numbers of steps which is linearly proportional to 
the number of distinct sites on the grain (Montroll \& Weiss, 1965) . 
If the grain having total $S$ number of sites then it should be proportional to $S$.
In reality, $S$ should be replaced by $S^\alpha(t)$, where
$\alpha(t)$ depends on various physical and chemical properties of grain (Chakrabarti et
al., 2006ab, Das et al., 2008). In Rate equation, this linearity constant ($\alpha(t))$ is taken to be one
but in actual case this may not be the case. Rate equation method is very popular because it is
economical  to run in computer, whereas Monte Carlo method takes a long computational time even to handle
a small chemical network. Chakrabarti et al. (2006ab) first proposed the idea for calculating
various values of $\alpha(t)$ to modify Rate equation in such a way that that one could have an
educated estimation of real scenario by using simple Rate equation method as well.
Chakrabarti et al. (2006ab) studied formation of $H_2$ on interstellar grains and proposed
values of $\alpha(t)$ under various physical circumstances. Das et al. (2008)
studied the formation of Water and Methanol around the dense cloud following the same approach. 
In both the cases, it was assumed that a steady state could be reached at 
$t\rightarrow \infty$ and at steady state $\alpha(t)=\alpha$. 
Here, we consider diffused cloud condition, where $H$ and $D$ are randomly accreting and
producing $H_2, \ D_2$ and $HD$. Due to inherent nature of the diffuse cloud condition,
steady state condition may never be achieved (or achieved during very late stage of evolution). 
So a steady state value of $\alpha$ could be misleading. Here, to avoid any discrepancies, 
we calculate a factor and named it scaling factor ($S_f$) defined to be the following:
$$
{\scriptsize S_f = \frac{Number \ of \ surface \ species \ X \ at \ time \ t \ by \ Monte \ Carlo \ method}
 {Number \ of \ surface \ species \ X \ at \ time \ t \ by \ Rate \ equation \ method}}
$$
This factor defines number of surface species as predicted by Monte Carlo method with respect that obtained by
Rate equation method. So after using Rate equation method, if we multiply the outcome by a
scaling factor ($S_f$) defined above, we may have a results as predicted by Monte Carlo method. 
In Fig. 6, we  show variation of 
$S_f$ for $H_2$, $D_2$ and $HD$ for an Olivine grain kept 
at $9\ K$, $n_H=10^4$ $cm^{-3}$, $S=4 \times 10^4$ with
number density of the cloud. For better illustration purpose, ${S_f}_{D_2}$ is multiplied by 
$10$ and ${S_f}_{HD}$ is multiplied by $5$. It is interesting to see that as we increase 
number density, $S_f$ is increasing. Physical significance is that as we are going to higher 
density regime, grain surfaces are more and more populated and the surface species could easily 
find its reactant partner to react, so the production is enhanced.

For low density case, production could be delayed due to unavailability of suitable reactant partner. 
Around high density region, production efficiency increases as well as $S_f$ increases. 
In Table 2, we present $S_f$ for all species ($H$, $D$, $H_2$, $D_2$ and $HD$) 
for various sets of binding energies at 
temperatures where their production efficiency is maximum.  For Amorphous Carbon grain, we provide
results for $T=15\ K$ and for intermediate energy values (set 3 energy values), $T=11\ K$. 
Similar trends could be observed for all sets of energies.

\begin {figure}
\centering{
\includegraphics[height=10cm,width=15cm]{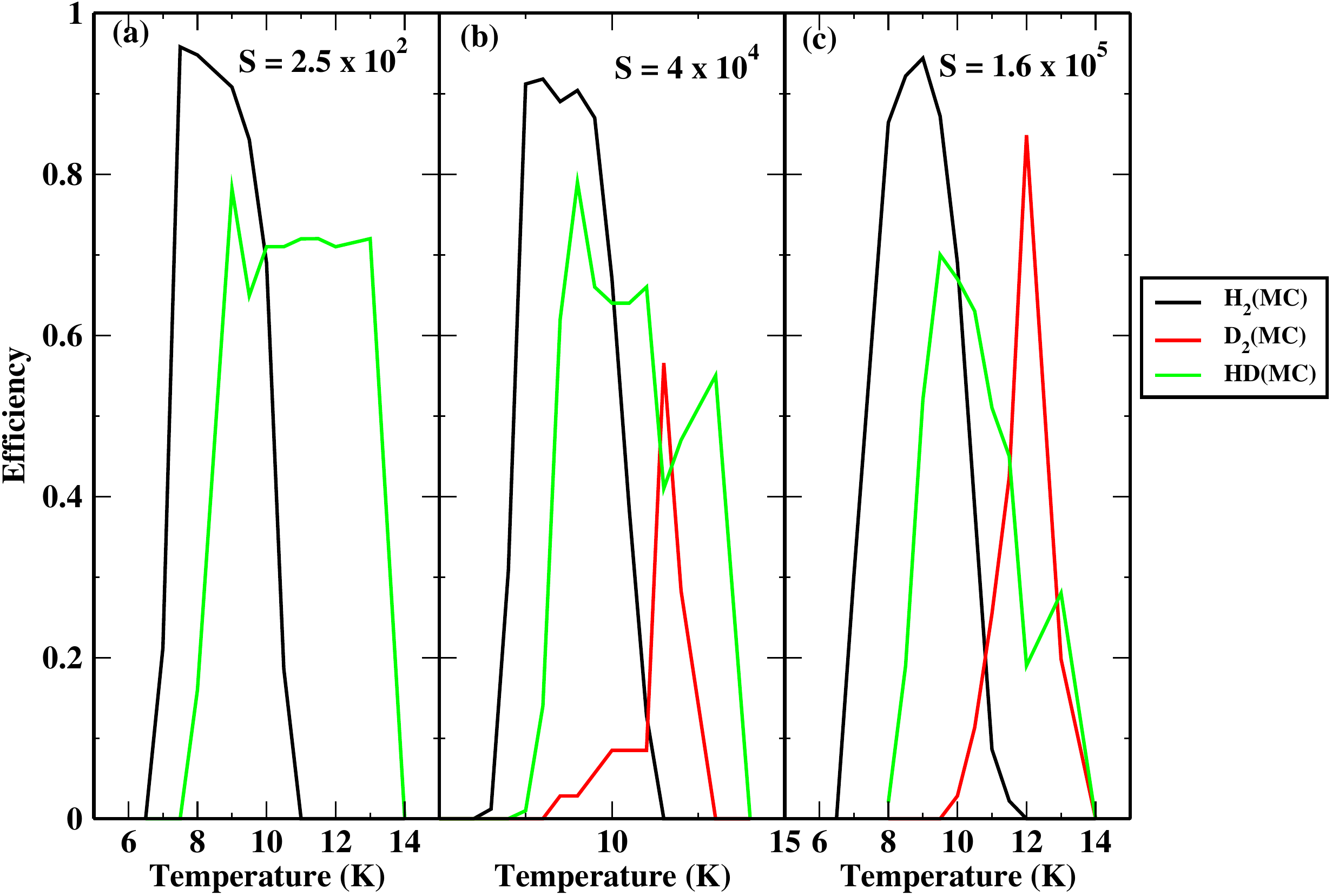}
\caption{Efficiency window of a Olivine grain having (a) $2.5 \times 10^2$, (b) $4 \times 10^4$ and
(c) $1.6 \times 10^5$ number of sites.}}
\label{fig-5}
\end {figure}

\begin {figure}
\centering{
\includegraphics[height=10cm,width=10cm]{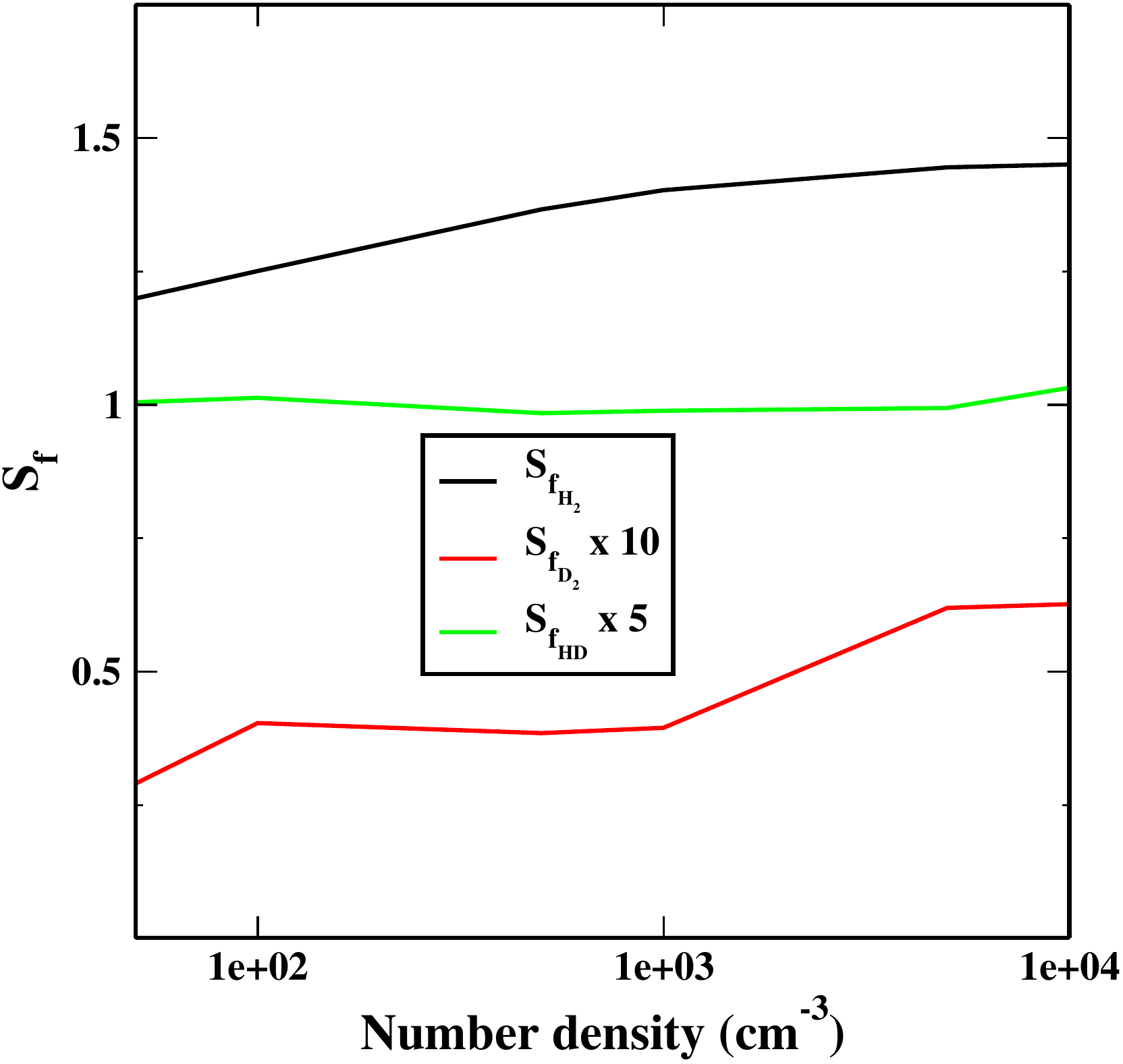}
\caption{Variation of $S_f$ of $H_2$, $D_2$ and $HD$ with number density of the cloud for an 
Olivine grain kept at $9\ K$. For the better visibility of the figure, ${S_f}_{D_2}$ and ${S_f}_{HD}$ are
multiplied by $10$ and $5$ respectively.}
\label{fig-6}}
\end {figure}

\begin{table}
{\centering
\scriptsize
\caption{Values of $S_f$ for a wide parameter space.}
\begin{tabular}{|c|c|c|c|c|c|c|c|}
\hline
Type of grain&Temperature (K)&Hydrogen number density (cm$^{-3}$) & ${S_f}_{H}$ & ${S_f}_{H_2}$ & ${S_f}_{D}$&${S_f}_{D_2}$&${S_f}_{HD}$\\
\hline
\hline
&&$50$&1.88 & 1.20&  0.25&  0.029&  0.20\\
Olivine&$9$&$100$&1.84&  1.25 & 0.24&  0.04&  0.20\\
&&$500$&1.86 & 1.37 & 0.245&  0.04 & 0.20\\
&&$1000$&2.35&  1.40 & 0.30&  0.04&  0.20\\
&&$5000$&0.84&  1.45&  0.06 & 0.06&  0.20\\
&&$10000$&1.21&  1.45&  0.09 & 0.06 & 0.21\\
\hline
&&$50$&1.78 & 1.41 & 0.22 & 0.03&  0.19\\
Amorphous Carbon&$15$&$100$&1.79 & 1.41&  0.22&  0.03 & 0.19\\
&&$500$&5.62 & 1.45&  0.37 & 0.06&  0.19\\
&&$1000$&0.28 & 1.46&  0.02 & 0.06& 0.19\\
&&$5000$&0.78 & 1.50&  0.06& 0.06&  0.20\\
&&$10000$&0.63 & 1.50&  0.045 & 0.07&  0.20\\
\hline
&&$50$&1.96&  0.91 & 0.29&  0.04&  0.20\\
Intermediate&$14$&$100$&1.95 & 1.07&  0.29 & 0.03 & 0.21\\
&&$500$&1.93&  1.30 & 0.27&  0.04 & 0.21\\
&&$1000$&1.99 & 1.34 & 0.28 & 0.04 & 0.21\\
&&$5000$&1.53&  1.38&  1.15& 0.05 & 0.203\\
&&$10000$&0.76&  1.41&  0.06& 0.07&  0.20\\
\hline
\end{tabular}}
\end{table}

\begin {figure}
\centering{
\includegraphics[height=10cm,width=10cm]{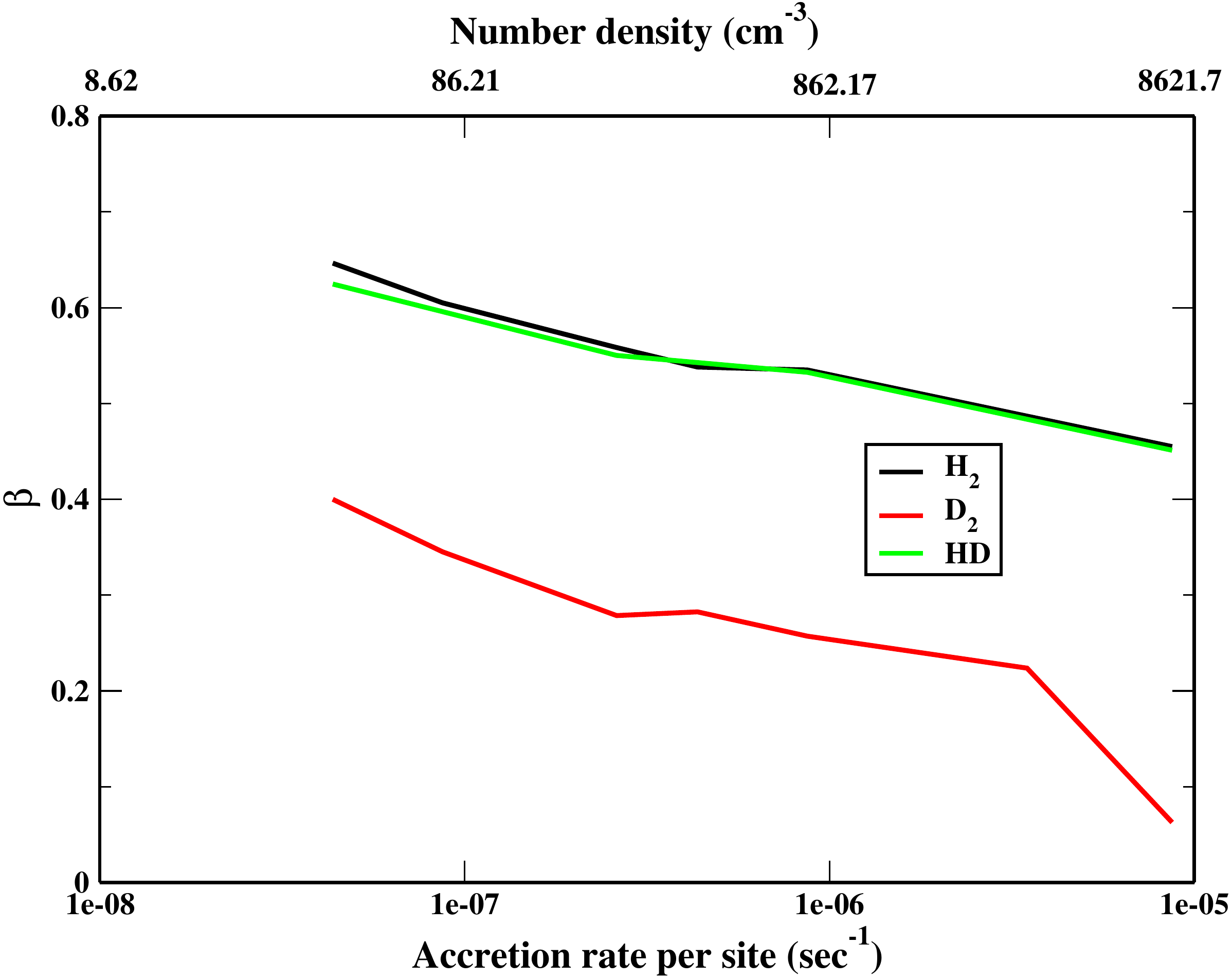}
\caption{Variation of $\beta$ with the accretion rate per site (lower label of X axis) 
or number density of the cloud (upper label of the X axis).}
\label{fig-7}}
\end {figure}

In order to see the effects of the physical parameters on the formation of H$_2$, HD and D$_2$,
we define another parameter $\beta$ and named it catalytic capability ($\beta$) by following
Chakrabarti et al. (2006ab) and  Das et al. (2008), which measures efficiency of 
formation of $H_2$, $HD$ and $D_2$ on grain surface for a given pair of species residing on it.
If $\delta N_{D_2}$ be the number of $D_2$ formed in $\delta t$
time then average rate of creation of $D_2$ per pair of deuterium atom is
given by,
\begin {equation}
\left< A_{D1}(t) \right > =\frac{1}{2N_D}\frac{\delta N_{D_2}}{\delta t} .
\end{equation}
We identify inverse of this rate as average formation rate and it is given by,
\begin{equation}
 T_f(t)=S^{\beta_{D_2}^{(t)}}/A_D .
\end{equation}
Thus,
\begin{equation}
S^{\beta_{D_2}(t)}=A_D/<A_{D1}(t)>
\end{equation}
This yields $\beta$ as a function of time:
\begin{equation}
\beta_{D_2}(t)=log(A_D/<A_{D1}(t)>)/log(S)
\end{equation}
Similarly for $H_2$ and $HD$,
\begin{equation}
\beta_{H_2}(t)=log(A_H/<A_{H1}(t)>)/log(S),
\end{equation}
\begin{equation}
\beta_{HD}(t)=log(A_{HD}/<A_{HD1}(t)>)/log(S),
\end{equation}
where, 
$$
\left< A_{H1}(t) \right > =\frac{1}{2N_H}\frac{\delta N_{H_2}}{\delta t}
$$
and
$$
\left< A_{HD1}(t) \right > =\frac{1}{(N_H+N_D)}\frac{\delta N_{HD}}{\delta t}.
$$
$\beta(t)$ is also a time dependent parameter. At $t \rightarrow \infty$, $\beta(t)\rightarrow\beta$. 
As time evolves, grains are populated by the species. Since, surface coverage 
is increasing, production should be faster due to decrease of reaction 
zone. But there should also be a blocking effect (Das et al. 2008), due to which surface species
could be locked for a while, which in turn could delay production process. 
So, value of $\beta(t)$ depends on surface coverage as well. 
After some transient time steps, we are having a steady state value of $\beta(t)$.
Here, we are considering at $t\rightarrow \infty$, $\beta(t) \rightarrow \beta$. 
In Fig. 7, we show 
the value of $\beta$ with respect to the accretion rate per site.  
$\beta$ values are calculated during the last few steps of our simulations ($\sim 10^8$ $sec$).
As expected, with increasing accretion rate, 
surface coverage of species increases. In this situation, 
surface species needs to travel lesser number of steps
to find one suitable reactant partner. Low value of $\beta$ thus signifies 
faster production whereas higher value represents slower production rate.

\begin {figure}
\centering{
\includegraphics[height=10cm,width=10cm]{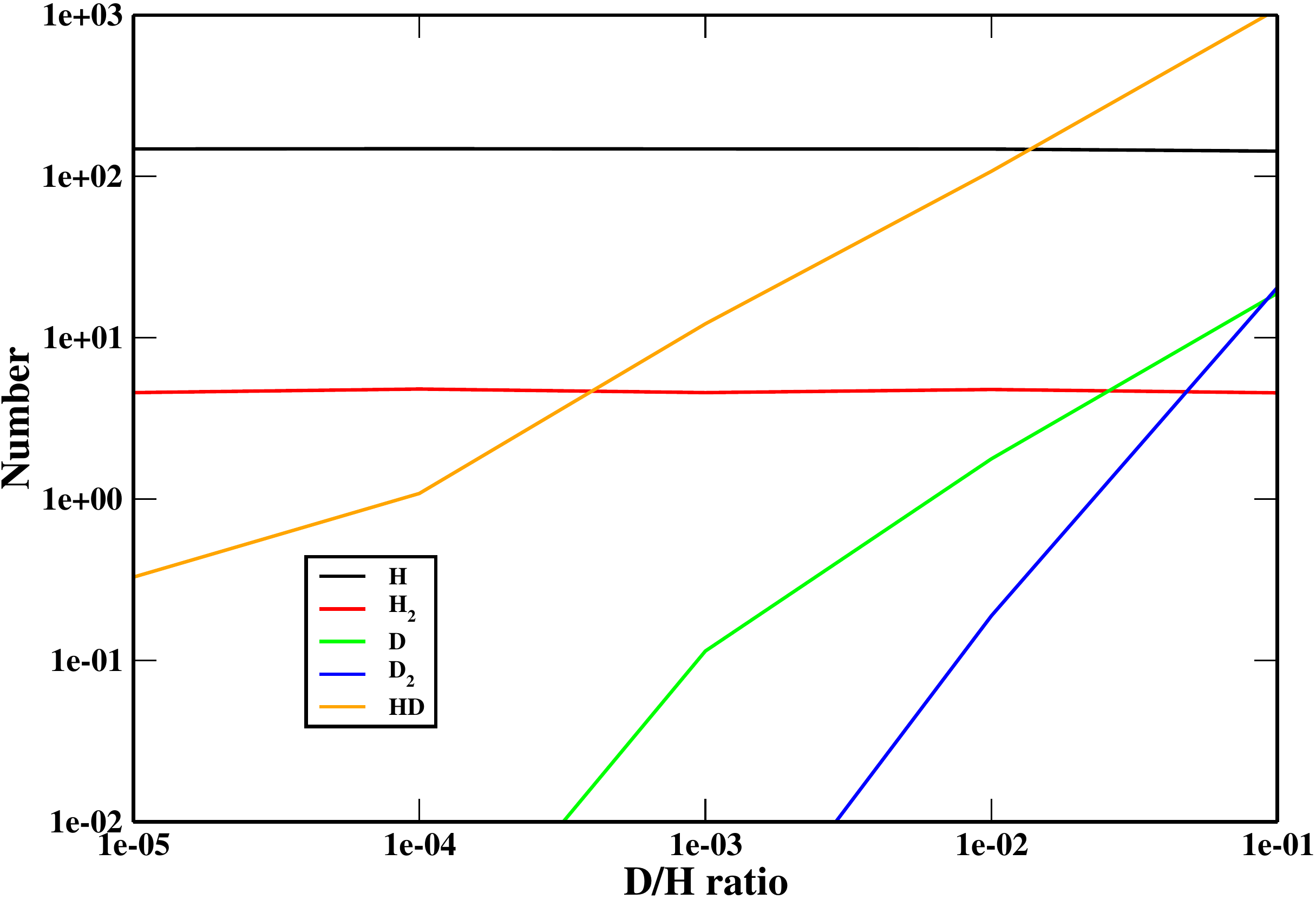}
\caption{Variation of the number of various surface species with the variation $r_D$.}
\label{fig-8}}
\end {figure}

In Fig. 8, number of various surface species are shown with variation of $r_D$. Observational
evidences suggest that elemental atomic $D/H$ ratio ($r_D$) in an ISM would be $\sim∼1.5 \times 10^{-5}$ 
(Linsky et al., 1995). Here, we consider an Olivine grain having
$4 \times 10^4$ sites kept at $T=9\ K$, $n_H=10^2$ cm$^{-3}$ and vary $r_D$ from 
$10^{-5}$ to $0.1$.  Due to randomness of Monte Carlo method, here, we 
plot average numbers (average taken from last few steps after $\sim 10^8$ $sec$).  
For lower value of $r_D$, production of $D_2$ is insignificant. Abundance of $H$ and 
$H_2$ remain almost constant throughout simulation range of $r_D$.

\section{Conclusion}

In this paper, we mainly focused on production of $H_2$, $D_2$ and $HD$ on grain surfaces.
Following are highlights of our results:

$\bullet$ {Production of these species highly dependent on the type of grains (i.e., interaction
energies). We carried out our simulations using three types of binding energies.
For Olivine grain, efficiency window is in between $8\ K-14\ K$, for amorphous carbon grain it 
is in between $11\ K-22\ K$ and for binding energies as in LBH (set 3 energy values), 
this window is shifted to $11\ K-19\ K$.}

$\bullet$ {We define a parameter $S_f$, which solely depend on various physical and chemical properties
of interstellar grains. Rate equation method often over or under estimates the production efficiency.
To obtain more accurate production of these simple yet the most abundant molecules
around ISM, this correction term should be considered in rectifying results from Rate equation method. For the
sake of wider usage of our parameters,  we provided a Table (Table 2) with various 
values of $S_f$ for a range of physical parameters.}

$\bullet$ {We computed another quantity $\beta$, named catalytic capacity as used in earlier papers.
This parameter shows a decreasing trend with increase in accretion rate. This implies
production of surface species more and more favourable for high accretion regime.}

$\bullet$ {Despite of low elemental abundances of atomic deuterium, several complex species are found 
to be heavily fractionated. Here, we vary initial $D/H$ ratio to find out deuterium fractionation of
simple yet the most abundant species, namely, $H_2$. If we consider  elemental $D/H$ ratio of $10^{-5}$
(Linsky et al., 1995) for a diffused cloud where all $H$ are in atomic form, production of $D_2$ found to be
insignificant.}

\section {Acknowledgment}
A.D. and D.S. wants to thank the ISRO respond project (Grant No. ISRO/RES/2/372/11-12) for financial support. 
L.M. is grateful to a MOES project for financial support.

\end{document}